\documentstyle[12pt]{article}
\input amssym.def
\input amssym
\input psfig

\newcommand{\reals}{\Bbb{R}}
\newcommand{\sphere}{\Bbb{S}}

\title{Computational Geometry Column 34}
\author{Pankaj K. Agarwal\thanks{
	Center for Geometric Computing,
	Dept. Computer Science,
	Duke University,
	Durham, NC 27708-0129. USA.
	{\tt pankaj@cs.duke.edu}.  
	}
\ and\ Joseph O'Rourke\thanks{
	Dept.\ of\ Computer\ Science,
	Smith College,
	Northampton, MA 01063,
	USA.
	{\tt orourke@cs.smith.edu}.
	}
}
\date{}
\begin{document}
\bibliographystyle{alpha}
\maketitle
\pagestyle{empty}
\thispagestyle{empty}

\begin{abstract}
Problems presented at the open-problem session of the
14th Annual ACM Symposium on Computational Geometry
are listed.
\end{abstract}

\begin{description}
\item[Jack Snoeyink, University of British Columbia:]\mbox{}
\begin{enumerate}
\item Given a set $P$ of points and a set $S$ of disjoint 
line segments in the plane, 
does there always exist a spanning tree of $P$ that, when embedded with 
straight edges, has the property that no segment in $S$ is cut 
by more than two edges?

\item If the weight of an edge of the spanning tree is the 
number of segments of $S$ that it crosses, 
then does the minimum spanning tree have weight at most $2 |S|$?  

One can ask the same questions for paths instead of trees.
\end{enumerate}

Note that an affirmative answer for the first problem implies 
an affirmative answer for the second.
While a few constructions give spanning trees with $O(|S| \log |S|)$ 
weight, the best known lower bound is  $2 |S|$.  
One can assume that the segments  of
$S$ induce a triangulation; 
in general, the application is to locate many points 
in a triangulation by walking along some ``nice'' path.
Disjointness of segments is important; otherwise spanning trees with low 
stabbing number give upper and lower bounds of $\Theta(|S| \sqrt{|S|} )$.
\item[Vadim Shapiro, University of Wisconsin:]
Given two smooth real algebraic hypersurfaces $S_1$ and $S_2$ defined by
polynomials of degree at most $d$ that are tangent along the curve 
$C$ (i.e., the locus of second order contact is  the curve $C$),
is there a (smooth?) algebraic hypersurface $S$ that is:
\begin{enumerate}
\item
 the zero set of a polynomial of degree at most $d/2$,
\item
 contains $C$, and
\item
 at each point of $C$,  $S$ meets both $S_1$ and $S_2$ transversally.
\end{enumerate}

If the degree of $S$ must be greater than $d/2$, 
what is the lowest possible degree of $S$ 
satisfying the other requirements?     

This problem arises in construction of Boolean  set 
representations for curved polyhedra;  the practical algorithms for 
constructing such surfaces are of particular interest.   
More details are in  [SV].

\begin{itemize}
        \item[{[SV]}]
V. Shapiro and D. L. Vossler,
        Separation for boundary to {CSG} conversion,
        \textit{ACM Transactions on Graphics\/}
        {\bf 12} (1993), 35--55.
\end{itemize}

\item[Joseph O'Rourke, Smith College:]

The cover of
{\em G\"{o}del, Escher, Bach} (by D.~R.~Hofstadter)
shows a solid piece of carved wood, which casts the letters
{\tt G E B}
as shadows in three orthogonal directions.
This suggests two questions:
\begin{enumerate}
\item Are there simple conditions on three shapes to be
realizable as shadows of a single, connected solid object
in $3$-space?
An example of realizable shapes is shown in Figure~1.
An example of unrealizable shapes is provided by the three letters
{\tt X X O}.
Are there any interesting classes of shapes such that any
three in the class are mutually realizable?
\item Extend this question to $d$ dimensions.
For example, can the letters
{\tt J O S E P H} be realized as the six $2$-dimensional 
shadows of a $4$-dimensional object?
\end{enumerate}
\begin{figure}[htbp]
\begin{center}
\ \psfig{figure=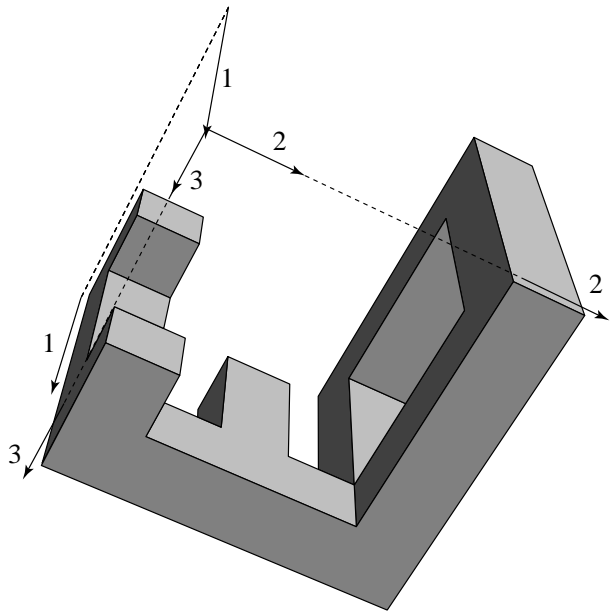,width=8cm}
\end{center}
\caption{An orthogonal polyhedron whose shadow in each of the three
labeled directions is an orthogonally polygonal letter of the alphabet.
Fig.~4.22 in~[OR].}
\end{figure}
\begin{itemize}
\item[{[OR]}]
J. O'Rourke, \textit{Computational Geometry in {C} (Second Edition)},
Cambridge University Press, Cambridge, 1998.
\end{itemize}

\item[Micha Sharir, Tel Aviv University:]
Let $S$ be a set of $n$ points in the plane, each moving with
a fixed velocity. 

How many times does the combinatorial
structure of the Euclidean Voronoi diagram of $S$ change over time?

The best known lower bound is quadratic, and the best known
upper bound is cubic; see [SA, Sec.~8.6.3]. If the distance is measured in
the $L_1$-metric, then the structure of the Voronoi diagram
changes only $O(n^2 \alpha (n))$ times [Ch], where $\alpha (\cdot)$
is the inverse Ackermann function.
\begin{itemize}
\item[{[Ch]}]
L. P. Chew, Near-quadratic bounds for the $L_1$ Voronoi diagram
of moving points, \emph{Proc.\ 5th Canad. Conf. Comput.
Geom.}, 1993, pp.\ 364--369.

\item[{[SA]}]
M.~Sharir and P.~K. Agarwal, {\em {Davenport}-{Schinzel}
Sequences and Their Geometric Applications}, 
Cambridge University Press, New York, NY, 1995.
\end{itemize}

\item[Subhash Suri, Washington University:]
Let $S$ be a set of $n$ points in the plane.  For a given
point $c$ in the plane, the \emph{star} of
$S$ is the set of edges $\{ cp \mid p \in S \}$.
The \emph{weight} of
a star  is the sum of the lengths of its edges. Let $MS (S)$
denote the minimum-weight star of $S$, where the minimum is
taken over all points in the plane, and let $MWT (S)$
denote the maximum-weight Euclidean matching of $S$. 

Obtain an upper bound on the ratio $MS (S)/MWT (S)$. 

It can be proved that $MS(S)/MWT(S) \le 2$, but the conjecture
is that the ratio is bounded by $2/\sqrt{3}$. The latter bound
can be achieved by many examples.

This problem arises in analyzing the performance of a 
network design heuristic. $MS(S)$ and $MWT(S)$ provide upper
and lower bounds, respectively, on the cost of traffic handling.
See [FST] for details.

\begin{itemize}
\item[{[FST]}]
A.~Fingerhut, S. Suri,  and J.~S.~Turner,
Designing minimum cost nonblocking communication networks,
{\em J. Algorithms\/} {\bf 24} (1997), 287--309.
\end{itemize}

\item[John Sullivan, University of Illinois:]
A knot $K$ in $\reals^3$ can be represented by its projection on
the $xy$-plane. The $xy$-projection is a self-intersecting
curve $K^*$.  At each such intersection point of $K^*$, 
we describe which of the two corresponding portions of $K$ lies 
above in the $z$-direction.
Given such a combinatorial description of  a knot  $K$ with $n$
intersection points in $K^*$, what is the minimum 
length of a rope of width $1$ that can realize $K$. 

Some knots may require the length of the rope to be $\Omega
(n)$, but is this an upper bound?  
It can be achieved with $O(n^2)$.

\item[Julien Basch, Stanford University:]

Let $S$ be a set of $n$ segments in the plane. For a given
$x$-coordinate $t$, let $\ell_t$ be the line $x=t$, and let 
$S_t \subseteq S$ be the set of segments intersected
by the line $\ell_t$. Suppose we construct a heap on
$S_t$, using the $y$-coordinates of the intersection points
of $S_t$ with $\ell_t$ as the keys. We maintain this heap as the
value of $t$ varies continuously from $-\infty$ to $+\infty$. 
That is, whenever $\ell_t$
 passes through an endpoint, the associated segment
is inserted into or deleted from the heap in the usual way, and
when $\ell_t$
passes an intersection point between two segments $p,q \in
S_t$ so that $p$
is the parent of $q$, the two elements are swapped to maintain
the heap property.

Obtain an upper bound on the number of swaps required 
to maintain the heap.

If $S$ is a set of lines, the best known upper bound on the 
number of swaps is $O(n \log^2 n)$.
For segments, the best known upper bound is $O(n^{3/2}\log n)$~[BGR].
\begin{itemize}
\item[{[BGR]}]
J. Basch, L. Guibas, and G. Ramkumar,
Sweeping lines and line segments with a heap,
\emph{Proc.\ 13th Annu. ACM Symp. Comput. Geom.}, 1997, pp.\ 469--472.
\end{itemize}

\item[Victor Milenkovic, University of Miami:]
Let $P$ be a simple polygon in the plane. Let $o$ be a
reference point in $P$, which coincides with the origin in the
standard placement of $P$. Let $\rho$ be a ray emanting from
$o$ attached to $P$. Let $P(\theta)$ denote the placement of
$P$ at which $o$ lies at the origin and the angle between the
$\rho$ and the $x$-axis is $\theta$.
Given a function $t : [\alpha, \beta] \rightarrow \reals^2$,
where $0 \le \alpha \le \beta \le 2\pi$, define
\[ \mathcal{I} (\alpha, \beta) = \bigcap_{\alpha \le \theta \le \beta}
	P(\theta) + t(\theta) .\]
Given $\alpha, \beta$, find a function $t(\theta)$ so that the
area of $\mathcal{I}(\alpha, \beta)$ is maximized.
Can $t(\theta)$ be computed efficiently?

What if we impose certain restrictions on $t(\theta)$? For
example, $t(\theta)$ can be restricted to be rotation with
respect to a point in $P$.

This problem arises in packing a family of polygons inside
another polygon; see~[Mi].

\begin{itemize}
\item[{[Mi]}]
V. J. Milenkovic,
Rotational polygon containment and minimum enclosure,
\emph{Proc.\ 14th Annual ACM Symp. Comput. Geom.}, 1998, 1--8.
\end{itemize}

\item[Joseph Mitchell, SUNY Stony Brook:] 
Let $S$ be a set of points in convex position in $\reals^3$.
Can the convex hull of $S$ always be triangulated 
(partitioned into tetrahedra)
so that the
dual graph of the triangulation has a Hamiltonian path?
The same question applies to points in convex position in
$\reals^d$, $d \ge 4$.

It is obviously true for points in convex position in the
plane.

Once such a triangulation is given for points in convex
position, then points can be added in the interior of the convex hull
and can be triangulated so that the new triangulation also
admits a Hamiltonian path.  See [AHMS].

\begin{itemize}
\item[{[AHMS]}] E.M. Arkin, M. Held, J.S.B. Mitchell, and S.S. Skiena,
Hamiltonian triangulations for fast rendering, {\em The
Visual Computer\/} {\bf 12} (1996), 429--444.
\end{itemize}

\item[Steve Vavasis, Cornell University:]
Let $P$ be a polyhedron in $\reals^3$, and let $\Sigma$ be a
simplicial complex.
Can one check in near linear time whether $\Sigma$ is a valid
mesh for $P$,  that is, whether the simplices in $\Sigma$ have
disjoint interiors and their union is $P$?

\item[Jack Snoeyink, University of British Columbia:]
Let $\Gamma$ be a set of $n$ $x$-monotone arcs with a total of $k$
intersection points. Every pair in $\Gamma$ intersects at
most once.  The following two operations are allowed on
$\Gamma$:
\begin{itemize}
\item[(i)] Given an arc, return its left and right endpoints.
\item[(ii)] Given an endpoint $p$ of an arc and another arc
$\gamma$, determine whether $p$ lies above or below $\gamma$,
or whether the $x$-projections of $p$ and $\gamma$ are
disjoint.
\end{itemize}
Using these primitives, 
how fast can one report all pairs of
intersecting arcs in $\Gamma$?

No subquadratic algorithm is known. 
An $\Omega(n\sqrt{k})$ lower bound is not difficult to prove.
Is there an algorithm that achieves this complexity?
A related question is how fast can one find a maximal set of
nonintersecting arcs.

This problem arises in developing an efficient, robust algorithm
for computing the intersection points of segments.

\item[Herbert Edelsbrunner, University of Illinois:]
Let $S$ be the set of vertices  of a strictly convex $n$-gon in
$\reals^2$. Let $u(S)$ be the number of pairs of vertices $p, q$ in $s$
at distance $1$ from each other.
Prove or disprove that $u(S) = O(n)$.

F\"{u}redi [Fu] proved that that $u(S)=O(n \log n)$, and
Edelsbrunner and Hajnal [EH] proved that there are strictly convex
$n$-gons in the plane for which $u(S) \ge 2n-7$.
\begin{itemize}
\item[{[EH]}]
H.~Edelsbrunner and P.~Hajnal, A lower bound on the number of
unit distances between the vertices of a convex polygon, 
{\em J. Combin.  Theory Ser.~A\/} {\bf 56} (1991), 312--316.

\item[{[Fu]}]
Z.~F{\"u}redi, The maximum number of unit distances in a
convex $n$-gon, {\em J. Combin. Theory Ser.~A\/} {\bf 55} (1990), 316--320.
\end{itemize}

\item[Herbert Edelsbrunner, University of Illinois:]
 An \emph{unfolding} of the boundary of a convex polytope $P$ is a
polygon obtained by cutting a tree on the surface that spans the
vertices.  The resulting flattened polygon has no polytope vertices
in its interior, and all interior ``fold lines,'' i.e., polytope edges,
leave no trace in the polygon.
However, we retain the gluing pattern by recording which pairs of
polygon edges were generated by a surface cut.  See Figure~2.

\begin{figure}[htbp]
\begin{center}
\ \psfig{figure=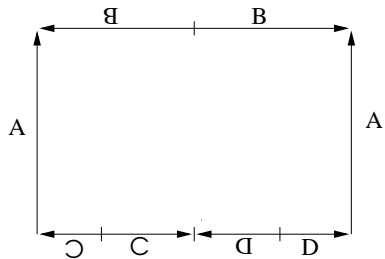,width=5cm}
\end{center}
\caption{Unfolding of a tetrahedron.}
\end{figure}

Alexandrov [A] showed that if the total angle is at most $2\pi$ at every
vertex and the gluing pattern results in a region homeomorphic to $\sphere^2$,
then the polygon is an unfolding of a unique convex polytope.

Give an algorithm that constructs the convex polytope from
the polygon.

\begin{itemize}
\item[{[A]}]
A. D. Aleksandrov, {\em Konvexe Polyeder}, Akademie Verlag, Berlin, 1958.
\end{itemize}

\item[Joseph Mitchell, SUNY Stony Brook:]
Let $S$ be a set of points in the plane so that the diameter
of $S$ is at most $2$. 

Is there a unit radius circle that
passes through exactly two points of $S$
(an {\em ordinary circle})?

The classic result of Sylvester shows that for $n$ points in
the plane, not all on a common line, there exists an {\em ordinary
line} (a line through exactly two points).  See~[BM,~CS] for a
summary of known results on this problem.
This question about unit
circles
[A. Bezdek, personal comm.]
is a natural generalization.

If diameter of $S$ is at most $\sqrt{2}$, then there is always
such a unit circle.

\begin{itemize}
\item[{[BM]}]
P. Borwein and W. Moser,  A survey of Sylvester's problem and
its generalizations,
\textit{Aequationes Mathematicae\/} {\bf 40} (1990), 111-135.

\item[{[CS]}]
J. Csima and E. T. Sawyer,
The $6n/13$ theorem revisited,
\textit{Graph Theory, combinatorics and Applications:
Proc. 7th Quadrennial Intl.\ Conf.\ Theory and
Appls.\ of Graphs}, vol. 1, (Y. Alavi and A. Schwenk, eds.), John Wiley
and Sons, Inc. 1995, pp.\ 235--249.
\end{itemize}

\item[Pankaj K. Agarwal, Duke University:]
Let $S$ be a set of segments in the plane. Suppose $S$
contains a subset $A$ of $k$ pairwise disjoint segments. 
Describe a polynomial-time algorithm to find a 
large subset of pairwise disjoint segments of $S$.

A few approximation algorithms are described in~[AKS,DMMMZ] if $S$ is
a set of orthogonal rectangles or circles.

\begin{itemize}
\item[{[AKS]}]
P. K. Agarwal, M. van Kreveld, and S. Suri,
Label placement by maximum independent set in rectangles,
to appear in \textit{Comput. Geom.: Theory and Appls.}

\item[{[DMMMZ]}]
S.~Doddi, M.~Marathe, A.~Mirzaian, B.~Moret, and B.~Zhu,
Map labeling and its generalization,
{\em Proc. 8th ACM-SIAM Sympos. Discrete Algorithms}, pp.\
  148--157, 1997.
\end{itemize}

\end{description}
\end{document}